# DAEMONS: DETECTION AT PULKOVO, GRAN SASSO, AND SOUDAN

E.M. DROBYSHEVSKI*, M.E. DROBYSHEVSKI, S.A. PONYAEV

*Ioffe Physical-Technical Institute, Russian Academy of Sciences, 194021 St-Petersburg, Russia*
*∗emdrob@mail.ioffe.ru*

I.S. GUSEVA

*Central Astronomical Observatory at Pulkovo, Russian Academy of Sciences,
196140 St-Petersburg, Russia*

During a week of the March maximum in 2011, two oppositely installed direction-sensitive TEU-167d Dark Electron Multipliers (DEMs) recorded a flux of daemons from the near-Earth almost circular heliocentric orbits (NEACHOs). The flux measured from above is $f \approx (8\pm3)\times10^{-7}$ cm$^{-2}$ s$^{-1}$, and that from below is twice smaller. The difference may be due both to specific design features of the TEUs themselves, and to dissimilarities in the slope of trajectories along which objects are coming from above or from below. It is shown that the daemon paradigm enables a quantitative interpretation of DAMA and CoGeNT experiments with no additional hypotheses. Both the experiments record a daemon flux of $f \sim 10^{-6}$ cm$^{-2}$ s$^{-1}$ from strongly elongated Earth-crossing heliocentric orbits (SEECHOs), predecessors of NEACHOs. Recommendations are given for processing of DAMA/LIBRA data, which unambiguously suggest that, in approximately half of cases (when there occur double events in the detector, rejected in processing under a single-hit criterion), the signals being recorded are successively excited by a single SEECHO object along a path of ~1 m, i.e., this is not a WIMP. It is noted that due regard to cascade events and pair interaction of ions will weaken the adverse influence exerted by the blocking effect on the channeling of iodine ions knocked out in NaI(Tl) crystal. This influence will become not so catastrophic as it follows from simplified semi-analytical models of the process: one might expect the energy of up to ~10% of primary recoil iodine ions will be converted to the scintillation light.

## 1. Introduction. Physics of Daemon Detection

Detection of daemons (DArk Electric Matter Objects), these elementary Planck black holes with a mass $m_{Pl} \approx 3\times10^{-5}$ g (radius $r_g \approx 2\times10^{-33}$ cm), which carry a negative electric charge of up to $Ze \approx 10e$ and constitute the dark matter (DM) of the Universe, is possible because they capture atomic nuclei with charge $Z_n$ and atomic mass $A_n$ when flying through a substance. The binding energy $W \approx 1.8ZZ_nA_n^{-1/3}$ MeV, released in this case, excites the nucleus, which leads to emission of atomic electrons, nucleons, clusters of these, and γ-photons [1]. These emissions are scintillation-active. Residing in the residue of a captured nucleus, a super-massive daemon carries-over this residue and small clusters of atoms (chemically) bound to this residue (we name such a complex as a c-daemon) in its motion through the substance and successively (at intervals of several microseconds) decomposes one nucleon after another in the nucleus. Modes of such a decomposition remain unclear [2], but, generally speaking, a sudden disappearance of each proton in the nucleus of an atom with $Z_n \approx 10\div30$ leads to energy release, with the resulting emission of a part of electrons of an atom (or a cluster), with an energy of ~1÷10 keV. In the end, the daemon frees itself of the residue of the nucleus and becomes capable of capturing a new one, with the whole process subsequently repeated.

The detector with a thin (~10 μm) ZnS(Ag) scintillating screen, developed on the basis of the above simplest ideas, enabled us to record a flux of these particles, which were first captured by the joint action of the Sun and the Earth from the Galactic disk in SEECHOs and then passed into NEACHOs and accumulated there, to subsequently fall on the Earth's surface at $V \approx 11.2\div15$ km/s.[2] The flux of NEACHO daemons varies with a period $P = 0.5$ yr and reaches the maximum values of $\geq10^{-7}$ cm$^{-2}$ s$^{-1}$ in March and September, which is understandable because many NEACHOs are nearly tangent to the Earth's orbit near the equinoxes [2]. It is noteworthy that the shape of the scintillation signal excited by a daemon capturing a nucleus in ZnS(Ag) is naturally characteristic of somewhat prolonged scintillations caused in ZnS(Ag) by heavy non-relativistic particles of the type of α-particles (heavy particle scintillations, HPSs). Joint analysis of results of many years' measurements shows that the confidence level of the March peak in the flux of NEACHO daemons presently reaches a value of 5σ [3].

Generally speaking, all physical experiments can be conditionally divided into two mutually overlapping groups: (1) reconnaissance experiments aimed to verify some working hypothesis, including approbation of a new device or approach in order to choose among several of these and to understand in which direction further studies are to be performed; and (2) routine exploratory experiments aimed to obtain statistically highly reliable results and be sure that the chosen way is correct, with the aim of a possible subsequent practical application of these results. As a rule, high measurement precision is not required for experiments of the first group, they only precede, and form the basis of, second-group experiments, being gradually transformed into these later.

Below are presented results of reconnaissance experiments aimed to reveal specific features of daemon detection by DEMs under conditions of various backgrounds and certain modifications of the recording system. These modifications improved the detection efficiency of NEACHO daemons (Secs. 2-4).

The suggestion about daemons falling down at a higher velocity ($V \approx$ 30-50 km/s) from intermediate SEECHOs quantitatively accounts [4] for the results and differences between the DAMA [5,6] and CoGeNT [7,8] experiments, and lack of their reproducibility in other experiments. This suggestion enabled predictions of some DAMA/LIBRA results, which were confirmed by subsequent experiments [9]. Now, on the same basis and without additional hypotheses, along with a quantitative interpretation of CoGeNT results, it is possible to give recommendations on further improving the efficiency of the DAMA/LIBRA detector (Sec. 5).

**2. Dark Electron Multiplier, a New Device for Detection of Daemons**

We found in our experiments that some of FEU-167 photomultiplier tubes we used [photocathode dia.100 (125) mm] can rather well detect by themselves, without an additional scintillator, flying through daemons. This property already follows from what was said above about mechanisms of interaction between daemons and a substance. It is apparent that, passing through the vacuum in a photomultiplier tube (PMT), atomic clusters carried-over by a c-daemon emit electrons, and some c-daemons (depending on their prehistory) can even completely free themselves of the residue of an earlier captured nucleus, so that, striking from the inside the wall of the PMT, they inevitably capture there a new nucleus, also with emission of electrons. The multiplication of electrons that have appeared in some way excites a signal at the PMT output, which has the shape characteristic of the intrinsic PMT noise (noise-like signals, NLSs). It is clear that the signal parameters will depend on specific features and properties of the inner coating and configuration of the PMT, etc. It is also apparent here that not all of these specific features affect the spectrophotometric properties of a PMT, and, therefore, they are not always controlled by manufacturers, being not necessarily identical even for PMTs from the same batch.

It was found that the highest sensitivity to daemons is exhibited by PMTs with a thicker inner aluminum coating of the near-cathode section of the tube (say, up to 0.5÷1-μm-thick, instead of the standard thickness of ~0.1 μm). This conclusion was supported by our reconnaissance underground experiments at Baksan, in which the triggering pulse was initiated by an HPS excited by a daemon in the ZnS(Ag) screen and the second, time-shifted pulse characterizing the daemon velocity was created by a light-insulated FEU-167 with increased thickness of the inner Al coating [19]. These experiments naturally suggested that the second, short pulse with an NLS shape was presumably created in our first experiments [1] in the lower FEU-167 itself, without the lower ZnS(Ag) scintillation screen being involved [3]. Therefore, a good reproducibility of these results would not be expected in replication of detectors of this kind with PMTs of other types (or even the same devices from other manufacturers) [3]. Actually, just only such a conclusion follows from measurements on the Nemesis installation [11] in which PMTs of 9265FLA and similar types with a 3" photocathode diameter (at a dynode section diameter of 51.5 mm) and a 49.4-mm-long near-cathode section were used. In FEU-167, these dimensions are, respectively, 5", 51.5 mm, and 65 mm, so that even geometrically, not to mention other possible differences, they must be more effective daemon detectors than 9265FLA. As a certain measure of the effective detecting area can serve, in the given case, the difference of the photocathode area and that of the projection of the cross-section of the cylindrical dynode section onto the photocathode; this parameter is 27 cm$^2$ for 9265FLA and 105 cm$^2$ (i.e., four times more) for FEU-167.

The finding that PMTs are sensitive to daemons naturally resulted in that their design was purposefully changed to make them sensitive to passing-through daemons and insensitive to photons or cosmic rays.

For this purpose, the manufacturer covered, on our request, the surface of the near-cathode section of the FEU-167 tube with a rather thick (~0.5 μm) Al layer. An experiment in March 2009 demonstrated that these DEMs (dark electron multipliers) do detect flight-through of daemons [3]. Further, the devices were improved so that they become sensitive to a direction of the daemon motion. For this purpose, a thick (~0.5 μm) Al layer was deposited only onto the inner surface of the planar front screen of FEU-167, whereas the side (cylindrical) and back (conical) surfaces of the near-cathode section of the tube were covered with a thin ~0.05 μm Al layer hardly transparent for visible light (its presence simultaneously ruled out generation of NLSs caused by scintillations excited in the tube glass by cosmic rays [12,13]).

Two such TEU-167d with horizontally oriented and nearly leaning against each other front screens were situated below a 3.5 mg/cm$^2$ ZnS(Ag) screen with an area of 0.5×0.5 m$^2$, viewed from above by an ordinary FEU-167, which generated HPS pulses triggering two digital two-trace S9-8 oscilloscopes (the intrinsic noise level in S9-8 is 0.2-0.4 mV, so that we reliably work with signals of ≥0.6 mV). The base distance between the ZnS(Ag) screen and the screens of TEU-167d was 29 cm. The experiment performed in September 2009 (although the latitude of St-Petersburg (SPb) is not the best place for observing the autumn maximum of the flux of NEACHO daemons) yielded positive results and unambiguously demonstrated that TEU-167d can detect daemons moving in a certain direction [14]. In data processing, we used the earlier discovered [1] dependence of the shape of HPSs on the motion direction of a daemon through a 3.5 mg/cm$^2$ ZnS(Ag) layer deposited onto a polystyrene substrate. Two these TEU-167d (nos. 0086 and 00105) with sockets containing the voltage dividers were transferred in February 2010 to the DAMA collaboration for usage in the Nemesis detector, with expectation of making experiments in March at more favorable conditions of the lower latitude of Rome. Unfortunately, our colleagues failed to immediately couple these devises with their electronics having a proper noise level of ~1 mV and a low input resistance (only 50 ohm, whereas the standard load of a PMT we exploited was some kiloohms), and could not return to this study later.

**3. DEM Experiment in Pulkovo**

DEMs were specially created for detection of daemons. Therefore, it is presently possible to refine and calibrate these devices by using fluxes of daemons themselves with more or less known (preliminarily determined) parameters. It is natural to take for this purpose the maximum March flux of NEACHO objects, whose existence reliability now exceeds 5$\sigma$ [3]. At a base distance of the detector of about 30 cm, an outfall of the NEACHO objects generates pulses spaced by ±(20÷30) μs.

Two TEU-167d newly fabricated to our order in Novosibirsk (no. 00160, channel 2; no. 00159, channel 21) were now mounted at the Central Astronomical Observatory of the RAS at Pulkovo, south of SPb, outside the city limits. We did so because, in contrast to the Ioffe Institute, Russian Academy of Sciences, the Pulkovo Observatory is not surrounded by multitude of research institutes and, therefore, the electromagnetic noise dramatically enhancing the electromagnetic background at the Ioffe Institute [3] must be not too strong there (the influence of the closely situated Pulkovo international airport onto our system was unclear).

We used the previously elaborated scheme [14] in which the HPS signal triggering two S9-8 oscilloscopes was provided by FEU-167 viewing from above a 22-cm ZnS(Ag) layer (FS-4 formulation) deposited onto the upper surface of a horizontal square polystyrene plate (0.5×0.5 $m^2$, thickness 3 mm). In order to raise the detection probability of daemons flying through the unit to the maximum possible extent, the thickness of the ZnS(Ag) layer was increased to ~7-10 mg/$cm^2$. For the same purpose, a voltage somewhat exceeding the optimal value (by ~50 V) was applied to both DEMs. Signals from the DEMs were delivered to the second traces of the oscilloscopes. A standard event was recorded with a personal computer if signals were presented on both traces of an oscilloscope. TEU-167d no. 00160 with FEU-167 viewing the ZnS(Ag) screen formed channel 2. It was oriented with its screen upwards, so that it had to predominantly record daemons moving through the Earth from below upwards. TEU-167d no. 00159 was situated in the middle of the module tin case. Together with FEU-167 viewing the ZnS(Ag) screen, it formed channel 21. The screen of this DEM was directed downwards, so that it had to record daemons predominantly moving from above downwards.

The system was exposed during a week from March 19 through 26, 2011 (actually six living days, with consideration for interruptions in its operation), when, according to our previous measurements [2,10] the flux intensity of NEACHO objects is close to the maximum value.

## 4. Experimental Results

The results of exposure of the module are shown in Figs. 1 and 2, with only events having a DEM signal amplitude in the range $1.0 \leq U_2 \leq 1.8$ mV represented. In accordance with expectations, channel 2, for which the front screen of the DEM faces upwards, shows in the distribution $N(\Delta t)$ of double events with a time $\Delta t$ shift (a total of 3365 NLS events from the TEU-167d, with amplitudes $1.0 \leq U_2 \leq 1.8$ mV) a maximum (360 events) in the bin $-40 \leq \Delta t \leq -20$ μs (Fig. 1a), which corresponds to existence of a flux of NEACHO daemons from below. By contrast, channel 21 (3227 events with $U_2 \geq 1.0$ mV, Fig. 1b) shows a maximum (386 events) in the bin $+20 \leq \Delta t \leq 40$ μs, which corresponds to a flux of NEACHO daemons from above downwards. The reliability of the -30 μs maximum for channel 2 is only $1.24\sigma$ (78.5%), and that for the +30 μs maximum for channel 21, $2.36\sigma$ (98%). Their summation gives $2.55\sigma$ (98.9%). At first sight, these values seem to be not very good, which was apparently caused by the fact that the background noise at Pulkovo, due to the proximity of the airport, still proved to be strong (indeed, the number of double events recorded by one channel during a week, $\sim 3 \times 10^3$, is rather large as compared with results of our previous experiments [1-3,10]). Nevertheless, for one-week experiment the C.L. $\approx 2.5\sigma$ is rather encouraging.

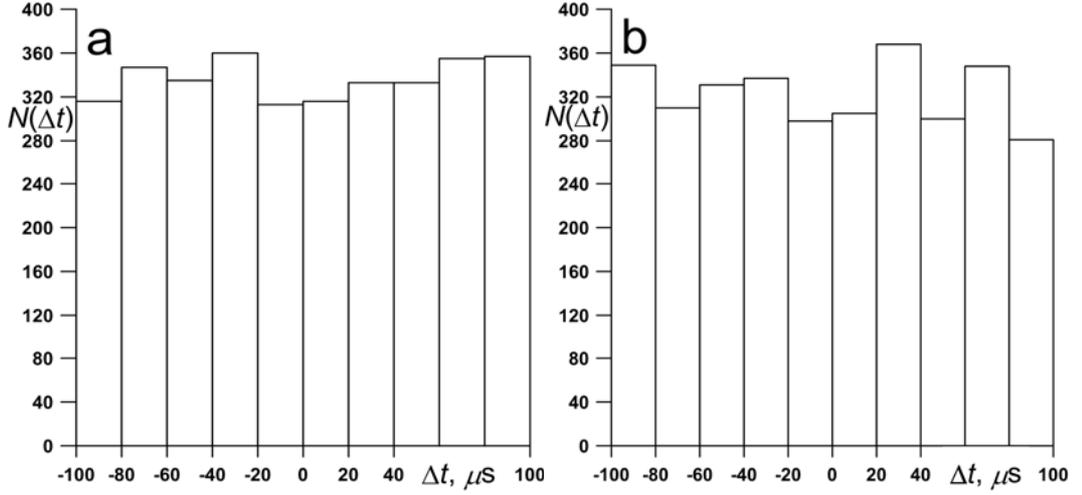

Fig. 1. Distributions $N(\Delta t)$ of double events with HPS from the triggering PMT over the time shift $\Delta t$ for channel 2 (Fig. 1a; TEU-167d no. 00160 is oriented with its screen looking upwards and, therefore, it must predominantly detect objects moving from below upwards, with $\Delta t < 0$, i.e., on flying through the Earth) and for channel 21 (Fig. 1b; TEU-167d no. 00159 has its screen looking downwards, it must detect objects incident from above, with $\Delta t > 0$). The maxima in the bins $20 < |\Delta t| < 40$ μs correspond to a vertical velocity component $V \approx 10 \div 15$ km/s. Number of events in sequential bins: (a) 316, 347, 335, 360, 313; 316, 333, 333, 355, 357; (b) 349, 310, 331, 337, 298; 305, 368, 300, 348, 281.

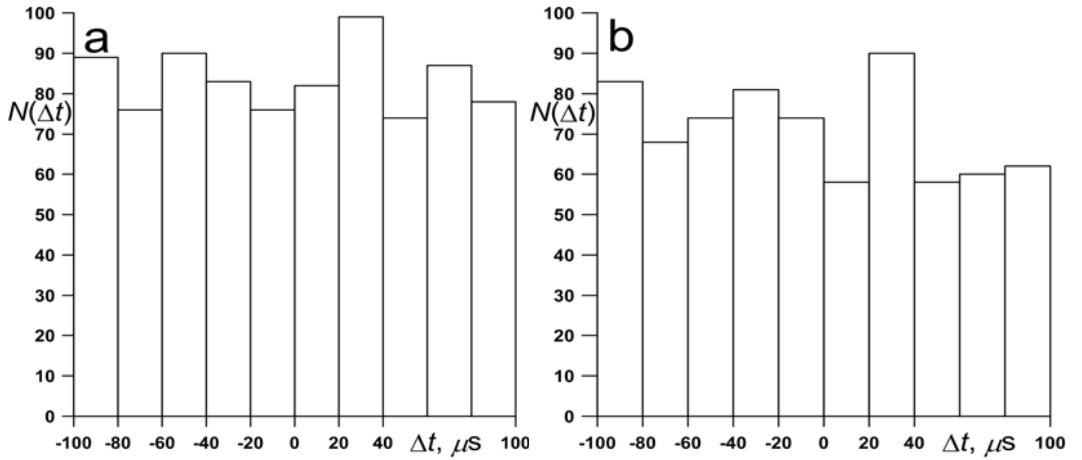

Fig. 2. Typical example of $N(\Delta t)$ distributions for double events with $U_2 = 1.0 \pm 0.1$ mV from TEU-167d no. 00159 (channel 21) for (a) HPSs narrower than the average value and (b) broader HPSs (the former are predominantly formed when a daemon flies from air through a ZnS(Ag) layer into a polystyrene substrate, and the latter, in a flight in the opposite direction [1]). It can be seen that, in the given case with a ZnS(Ag) coating with a thickness raised to $7 \div 10$ mg/cm$^2$, there is no noticeable dependence of $N(\Delta t)$ (attention should be given to the maximum at $20 < \Delta t < 40$ μs) on the HPS width. The average width of HPS pulses was determined for the whole set of events with $U_2 \geq 1$ mV.

The data obtained suggest the following:

(1) Channel 21 (TEU-167d no. 00159) records a nearly twice larger flux of objects (from above), compared with channel 2 (TEU-167d no. 00160). Among possible reasons are differences in the quality and sensitivity of instruments and purely geometrical properties of trajectories of daemons being detected. Channel 21 mostly records primary NEACHO objects falling onto the Earth from above downwards. In March, many of their trajectories in SPb are directed nearly perpendicularly to the Earth's surface [2,10]. Channel 2 records objects passing through the Earth and scattered by its gravitational field. Therefore, there appear trajectories nearly tangential to the Earth's surface. If the angle of a trajectory to the surface is ≤π/4, objects moving along such trajectories do not pass successively through the ZnS(Ag) scintillation screen and the DEM screen, i.e., these objects do not produce double time-shifted signals indicating that they pass through the detector.

(2) The primary flux of NEACHO daemons, measured during six living days from above through the screen of DEM no. 00160 (area 110 cm$^2$), reaches a value $f = (368-322.7)/(6\times86400\times110) \approx (8\pm3)\times10^{-7}$ cm$^{-2}$ s$^{-1}$. This value of the flux somewhat exceeds results of our previous measurements (up to $f \approx 3.4\times10^{-7}$ cm$^{-2}$ s$^{-1}$ [3]). As a possible reason for the increase in the detector efficiency can serve the above-mentioned two factors: (1) nearly optimal design of DEMs, together with the intended increase in their voltage; and (2) increase in the ZnS(Ag) layer thickness to 7-10 mg/cm$^2$. At the same time, it is known [13] that, beginning with a thickness of ~15 mg/cm$^2$, a layer of the ZnS(Ag) powder becomes nontransparent for the proper radiation, so that the detection efficiency falls. In addition, it should be noted that, even at a thickness of 7-10 mg/cm$^2$ of a ZnS(Ag) layer on polystyrene, the HPS shape ceases to be dependent on the direction in which NEACHO daemons move through this layer [see Fig. 2; in this case, the reliability of the +30 μs maximum is $2.53\sigma$ (98.8%)]. The ZnS(Ag) layer has a geometric thickness of ~20 μm. Hence follows that, for the effect to disappear, the de-excitation fime of a nucleus captured and carried-over by a NEACHO daemon ($V = 10 \div 15$ km/s) must be $\leq (1 \div 2) \times 10^{-9}$ s, in agreement with our previous estimates [1]. It is noteworthy that it should be kept in mind that the binder mass, in its turn, should be smaller than the ZnS(Ag) mass in an analysis of how the shape of HPSs depends on the direction in which daemons move through a ZnS(Ag) layer.

## 5. Some Consequences and Recommendations. DAMA and CoGeNT Experiments: Daemons or WIMPs?

It should be accepted that the main result of the present study is that the daemon paradigm has reached such a level that, based on this paradigm, we are not only capable of consistently explaining the few available positive results of a search for DM objects and prognosticating further evolution of this search, but also can create effectively operating novel devices on its basis.

### *5.1. Comments on a channeling efficiency*

Indeed, we have shown previously [4] that all the highly (at C.L. ~ $9\sigma$) reliable DAMA/NaI results obtained in underground experiments at Gran Sasso [unexpectedly low signal level: $\leq 6$ keVee (electron equivalent) instead of 10-50 keV conventionally expected from galactic WIMPs, the annual periodicity of the measured flux of objects, and the amplitude of its variation] are unambiguously and simply quantitatively(*sic*!) accounted for, without additional assumptions, by the fallout onto the Earth at $V = 30\text{-}50$ km/s of a flux (up to $f \approx 10^{-6}$ cm$^{-2}$ s$^{-1}$) of daemons captured in SEECHOs mostly oriented toward the solar antapex (this predetermines the annual periodicity of their flux). The energy of 6 keVee just corresponds to the energy of an iodine ion with $V \approx 100$ km/s, elastically knocked by a super-massive particle with $V \approx 50$ km/s. This, however, requires that the quenching factor $q$ for a certain fraction of ions knocked out from the crystal lattice of the scintillator should be close to unity ($q$ is the ratio between the scintillation intensity from an ion and an electron with the same energy). The requirement $q \to 1$ is satisfied for ions channeled in a crystal, when most part of their energy is transferred to the electronic component of the crystal (see ref. [4] and refs. therein). An important factor in our estimates of the flux intensity measured by DAMA and analyses of the correspondence between the scintillation energy and energy of DM objects (SEECHO daemons) was the conclusion about channeling of up to ~20% of iodine ions elastically knocked out by these objects in NaI(Tl) crystal.

However, Bozorgnia *et al* [15] draw attention to the circumstance that the fraction of channeled ions in the given case, when they are knocked out from equilibrium positions in the crystal, will be negligible because of their blocking by neighboring atoms (so-called "shadow effect") [16]. This blocking prevents the recoil ions from immediately entering the channel at angles necessary for the channeling conditions to be satisfied. According to Bozorgnia *et al* [15], the only apparent factor favoring capture of a recoil ion into the channel are thermal vibrations shifting the ion from the equilibrium position in the crystal, so that it is already in the channel itself when being knocked out. However, in this case too, only a small fraction of a percent of recoil iodine ions with a keV energy are channeled in NaI at room temperature.

Bozorgnia *et al* calculations were made in a quasi-analytical approximation going back to Lindhard's approach [17] in which the field potential in a crystal was approximated with rather smooth functions, and secondary processes of the type of scattering and multistage energy transfer to other ions, in which occurs their channeling or rechanneling of an already scattered primary ion, were not studied deeply owing to their complexity (see, however, ref. [18] and refs. therein). Thus, these pessimistic conclusions should be regarded as certain limiting estimates. Indeed, the authors of ref. [19] themselves note that, to take into account the multistage nature and diversity of the occurring processes, "montecarlo simulations may be needed to settle these issues".

It should be admitted that the relevant science is still insufficiently advanced. It is more or less developed and reliable for single-element crystals of the type Si, Ge [20], Xe, Ar, Ne [21], etc. A single-element approximation is used for KCl crystals on the assumption that the masses of K and Cl ions are the same and have a certain intermediate value. This is to a certain extent admissible for a crystal like CsI, with ions of nearly equal masses [22], but is hardly acceptable for NaI, in which these masses differ by a factor of 5.5. It should be kept in mind here that six ions closest to the heavy iodine ion (and just this ion created, in the case of its channeling, 6 keVee scintillations in DAMA/NaI) are light $Na^+$ ions. These latter can hardly create a full-rate blocking for a knocked-out iodine ion. At the same time, it seems that nobody has attempted to take into account the decrease in the blocking effect of small-mass neighboring ions that are also experiencing thermal vibrations with a frequency of $5.5^{1/2}$ times (for NaI) that for iodine ions, or, say, different collective vibrations and processes (of the type of the Mössbauer effect etc.). Finally, it seems that nobody has considered, either, the possibility and consequences of channeling of whole molecules, say, a NaI molecule formed by a recoil ion and an ion trying to block its channeling. Studies devoted to channeling in multiple-element crystals disregard processes in which chemical bonds continuously formed between a moving ion and nearest atoms are ruptured, whose importance was noted by Firsov [23] long ago. These processes convert the energy of valence electrons to emitted light more effectively than does irradiation of a crystal with keV electrons, so that it may occur that $q > 1$ for a certain fraction of channeled ions [4]. Thus, the above-mentioned factors act in the same direction: they weaken the blocking effect.

Taking into account the self-consistency of conclusions made about DAMA experiments (and now also about CoGeNT experiments, see below) in terms of the daemon paradigm and, if you wish, being guided by the requirement that a theory must be beautiful ("This result is too beautiful to be false", - utterance ascribed to P.A.M. Dirac; see also ref. [24]), we can well make an opposite assumption. Namely, into the scintillation light nearly fully passes (via excitation of the electronic component) the energy of ~10% of the whole amount of primary recoil iodine ions in a NaI(Tl) crystal (for this part of ions, $q \approx 1$).

*5.2. How to improve an efficiency of the DAMA/LIBRA detector*

Based on the daemon paradigm, we could demonstrate that the model-dependent single-hit criterion (SHIC) in which closely time-spaced pair events in different crystals of the detector assembly are excluded from consideration, fully justified when WIMP candidates for the role of DM objects are considered, should not be used for detection of daemons, particles that can be in no way regarded as weakly interacting [9]. It has been noted that application of SHIC to interpretation of data from more massive detector assemblies will impair their efficiency [25], i.e., will lead to loss of information because of the decrease in the number of events being recorded in unit time per unit detector mass. This prediction has been confirmed by DAMA/LIBRA experiments with 25 NaI(Tl) crystals, instead of 9 in DAMA/NaI. The detector efficiency per unit mass in 2-6 keVee range decreased nearly twice (the amplitude of annual variations of the registered object flux has dropped from $A = 0.020$ cpd/kg/keV to $A = 0.0107$ cpd/kg/keV) [6]. In the first publications on DAMA/LIBRA, the authors of the experiment tended to believe that the decrease is accidental and occurs for purely statistical reasons [6]. At the same time, data of recent years indicate that the decreased value $A = 0.0097\pm0.0015$ cpd/kg/keV for DAMA/LIBRA is reliable [26], i.e., the predicted effect is confirmed. Hence follows that application of SHIC, based on the model-dependent assumption that WIMPs are detected, is not justified and contradicts the experiment. In particular, the application of SHIC also accounts for the observed decrease in the upper level of signals (from 6 keVee in DAMA/NaI [5] to 5 keVee and even less in DAMA/LIBRA [6,25]) [9].

In a sense, the DAMA experiments with NaI(Tl) scintillators in its modern form have exhausted itself: obtaining greater statistics, without extracting a new physics, is useful up to a definite level, - indeed one does not feel a real difference between confidence levels of 9$\sigma$ or 19$\sigma$, say.

A highly important result can be obtained by demonstrating that the objects detected in DAMA experiments are not WIMPs. This can be done rather simply: it is only necessary to select in a 25-element DAMA/LIBRA assembly at least two compact assemblies with 9 elements (say, crystals 1, 2, 3, 6, 7, 8, 11, 12, 13 and 13, 14, 15, 18, 19, 20, 23, 24, 25) and to independently process their data by a receipt used in processing of DAMA/NaI results (even if using SHIC). One can be sure that, in this case, the efficiency of event detection (and, consequently, that of the DAMA/LIBRA detector as a whole) will increase: the amplitude *A* of the annual variation of fluxes of DM objects (daemons!) will double and will again reach the values *A* = 0.020 cpd/kg/keV that follow from the DAMA/NaI experiment [5]. Simultaneously, the mean free path between effective elastic interactions of objects being detected (c-daemons) with atomic nuclei of a substance will be estimated: it will constitute tens of centimeters, rather than light years (and the interaction cross-sections, large fractions of barns, rather than small fractions of a picobarn).

All the aforesaid refers to the future 1-ton NaI(Tl) DAMA experiment.

### 5.3. Daemons or WIMPs? CoGeNT and further on…

The insistent faith in WIMPs is surprising and admirable. During the last 20 years, the sensitivity of experiments devoted to search for these objects has been raised by a factor of $10^4$ to $10^5$ without any noticeable effect, which once again suggests the WIMP hypothesis is of *ad hoc* nature.

The recently reported positive results of CoGeNT at C.L. ≈ 2.8$\sigma$ [7,8] (this experiment is being carried out at the Soudan Underground Laboratory) can be matched with the highly reliable DAMA data only in terms of WIMP models with up to ten(!) free parameters (e.g., refs.

[27,28]). At the same time, it can be easily shown without additional assumptions and parameters that, similarly to DAMA, CoGeNT records a flux of SEECHO daemons incident on the Earth at $V \approx 30\text{-}50$ km/s. Indeed, an elastic interaction of such an electrically neutral c-daemon moving at $V \approx 50$ km/s with an iodine ion ($A_n = 127$) in a NaI(Tl) crystal imparts to this ion an energy of up to 6.6 keV, and that with a Ge ion ($A_n = 72.6$) in the CoGeNT crystal, an energy of up to 3.7 keV. However, just these values are characteristic of the upper limits to signals (6 keV in DAMA and ~3 keV in CoGeNT; in the latter case, the annual modulation of the flux is observed in the energy range 0.5-3 keV).

Simple estimates made in the gas-kinetic approximation yield for the flux of SEECHO daemons a value $f \approx 3 \times 10^{-7}$ cm$^{-2}$ s$^{-1}$ [20]. This value is in a rather good agreement with the experimentally measured flux of NEACHO daemons: up to $f \approx 8 \times 10^{-7}$ cm$^{-2}$ s$^{-1}$ (see above). DAMA/NaI experiments [5] (but not DAMA/LIBRA for which data processing with SHIC is impermissible, see above) give, with the channeling taken into account, $f \approx 6 \times 10^{-7}$ cm$^{-2}$ s$^{-1}$ during its summer maximum [4] (if $q \approx 1$ for ~10% of primary iodine recoil ions, $f$ increases to ~$10^{-6}$ cm$^{-2}$ s$^{-1}$, see above).

In the CoGeNT experiment, the crystal is ~5 cm in size (its cross-section area is ~25 cm$^2$), the number of events is as large as once a day (see Fig. 4 in ref. [8]). Hence follows that $f \approx 1/(86400 \times 25) \approx 5 \times 10^{-7}$ cm$^{-2}$ s$^{-1}$. A better coincidence with the above estimates in such a rough approach could hardly be expected, especially if we take into account that the free path of a neutral c-daemon between elastic interactions with nuclei exceeds the crystal size somewhat.

It is, of course, apparent that all these calculations need refinement and, among other things, require consideration of numerous fine and still not always clear mechanisms of interaction of daemons with a substance, including diverse (blocks of concrete, paraffin, lead, copper, etc.) environments of detecting elements in different systems.

It is hoped that all (positive) results to be furnished by other WIMP detectors after they reach the sensitivity necessary for recording elastic interactions of a c-daemon with atomic nuclei (~1÷5 keV, depending on the material of a detecting substance), will be equally easily interpreted and interrelated in terms of the daemon paradigm. At the same time, as it follows at least from the results of the present study, the daemon paradigm provides a key to understanding of how the experiments being already performed should be modified to be successful.

**Acknowledgments**

The authors are greatly indebted to A.S.Rybak, B.K.Ermakov, and Yu.S.Likh from the Pulkovo Observatory for help in carrying out the experiment.